# Trigger mechanism for a singing cavitating tip vortex

**Zhaohui Qian[1,2], Yongshun Zeng[1,2], Xiaoxing Peng[3] and Xianwu Luo[1,2,†]**

[1]Beijing Key Laboratory of $CO_2$ Utilization and Reduction Technology, Department of Energy and Power Engineering, Tsinghua University, Beijing 100084, China;

[2]State Key Laboratory of Hydro-science and Engineering, Tsinghua University, Beijing 100084, China;

[3]National Key Lab on Ship Vibration and Noise, China Ship Scientific Research Center, Wuxi 214082, China;

† Email address for correspondence: luoxw@tsinghua.edu.cn

The discrete tone radiated from tip vortex cavitation (TVC), known as 'vortex singing', was recognized in 1989, but its triggering remains unclear for over thirty years. In this study, the desinent cavitation number and viscous correction are applied to describe the dynamics of cavitation bubbles and the dispersion relation of cavity interfacial waves. The wavenumber-frequency spectrum of the cavity radius from the experiment in *CSSRC* indicates that singing waves predominantly consist of the stationary double helical modes ($k_\theta$ = $2^-$ and $-2^+$) and the breathing mode ($k_\theta$ = $0^-$), rather than standing waves as assumed in previous literatures. Moreover, two trigger mechanisms, expressed by two triggering lines, are proposed: the twisted TVC, initially at rest, is driven into motion through the corrected natural frequency ($f_n$) due to the step change of the far-field pressure. Subsequently, the frequency associated with the zero-group-velocity point ($f_{zgv}$) at $k_\theta$ = $0^-$ is excited through $f_i$, the frequency at the intersection of dispersion curves at $k_\theta$ = $0^-$ and $-2^+$, or $f_j$, the frequency at the intersection of dispersion curves at $k_\theta$ = $0^-$ and $2^-$, corresponding to two types of the vortex singing triggering. These solutions, without empirical parameters, are validated using singing conditions provided by *CSSRC* and *G.T.H.*, respectively. Furthermore, the coherence and the cross-power spectral density spectrum indicates a large-scale breathing wave propagating along the singing cavity surface and travelling from downstream to hydrofoil tip, providing us a comprehensive understanding for the triggering of vortex singing.

## 1. Introduction

Cavitation has been an attractive topic in the field of hydraulic machinery and ocean engineering for over one hundred years (Arndt *et al.* 1991; Brennen 2014; Luo *et al.* 2022). Among various types of cavitation, the tip vortex cavitation (TVC) is the first one to appear on propellers or axial-flow turbines, acting as a crucial source for underwater noise and vibrations (Posa *et al.* 2022; Qian *et al.* 2022; Ji *et al.* 2023). Consequently, many studies focused on forecasting TVC's inception (Arndt & Keller 1992; Amini *et al.* 2019; Chen *et al.* 2019), capturing general features such as the vortex trajectory, vapor core radius and roll-up kinematics (Asnaghi *et al.* 2020; Xie *et al.* 2021; Cheng *et al.* 2021; Russell *et al.* 2023), with few discussions concerning the cavity dynamics and noise generation mechanisms (Pennings *et al.* 2016a; Liu & Wang 2019; Klapwijk *et al.* 2022; Wang *et al.* 2023).

A representative noise enhancement phenomenon, excited by the TVC's interfacial instability, is the emission of a relatively higher amplitude and discrete tone from the fully developed TVC under certain operation conditions, which was firstly reported by Higuchi (1989) and named as 'vortex singing' by Maines & Arndt (1997). They tested different hydrofoils by varying the velocity, angle of attack and water quality in two facilities, one in *Obernach*, Germany and one at St. Anthony Falls Laboratory (*SAFL*), USA. The vortex singing was triggered within a distinct range of singing cavitation number, $\sigma_s$, and the standing wave along the cavity surface appeared to oscillate in phase with the sheet cavity at the tip. Furthermore, the singing frequency, $f_s$, typically changes from 400 *Hz* to 1.1 *kHz* and induces an intense peak of 20 dB to 30dB above the background in the noise spectrum (Arndt *et al.* 2015). Briançon-Marjollet and Merle (1996) described their observations on the singing vortex over the hydrofoil with a NACA 0020 cross-section in a larger cavitation tunnel, G.T.H., however the available conditions deviated from the narrow range of both $\sigma_s$ and $f_s$ defined for singing points by *SAFL* and *Obernach*.

One of the elliptical foil geometries, with NACA $66_2$-415 section, was tested in *Delft* University of Technology by Pennings *et al.* (2015a). Although the vortex singing did not appear in the experiment, the property of TVC dynamics agrees well with that in *SAFL* and *Obernach*. The same foil has been tested in a smaller cavitation tunnel, *CSSRC*, by Peng *et al.* (2017a). They developed a reliable experiment process to trigger the vortex singing by varying the cavitation number, $\sigma$, with a step change until it approaches $\sigma_s$. Song *et al.* (2018) also regenerated the TVC-singing in *CSSRC* and further pointed out that the inflow dissolved gas ($D_O$) imposed a significant and similar effect on both the singing cavitation number, $\sigma_s$, and the desinent cavitation number, $\sigma_d$. Recently, Ye *et al.* (2021, 2023) found the travelling of breathing waves along the TVC at Zhejiang University (*ZJU*), similar to the travelling waves of vortex singing at *CSSRC*. However, no vortex singing could be heard. Due to the higher dissolved gas content and $\sigma_d$ in their experiments, it could be inferred that the vortex singing is sensitive to the distance between $\sigma$ and $\sigma_d$. Because the triggering theories of vortex singing remains unclear, neither experiment nor numerical simulation can accurately excite the intense cavity resonance and the discrete emitted noise from a developed TVC (Simanto *et al.* 2023; Klapwijk *et al.* 2022).

To find the trigger mechanism for a vortex singing from TVC, two-dimensional (2-D) cavitation bubble kinematics and dynamics were examined theoretically and numerically (Choi & Ceccio 2007; Choi *et al.* 2009; Bosschers 2018a). The self-excited frequency $f_n$, known as the natural oscillation frequency, seems to be a primary resonance source for the singing frequency $f_s$. However, the 2-D analytical prediction with acceptable accuracy for $f_n$ becomes extremely challenging (Bosschers 2009b). Therefore, the three-dimensional (3-D) dispersion relation was further applied to solve the $f_n$. As suggested by Keller & Escudier (1980), the standing wave is possibly generated if the vortex is superimposed on the uniform axial flow. Using this idea, Maines & Arndt (1997) explained the determination of $\sigma_s$ and $f_s$ in *SAFL* and *Obernach*, but the agreement between their results and that in *G.T.H.* is poor and not acceptable. The dispersion relation for predicting deformations over a TVC was proposed by Pennings *et al.* (2015a, 2015b), applying the *Lamb-Oseen* vortex model. Based on the same consideration, Bosschers (2009a) introduced the semi-empirical viscous correction and left the criterion of zero-group-velocity point on breathing waves as the most likely hypothesis for stimulating $f_s$, with other criteria, for instance, the crossing points of dispersion lines for breathing waves and double helical waves (Bosschers 2018a). The main issue with his model is its inability to predict $\sigma_s$ for vortex singing (Pennings *et al.* 2015b). Furthermore, it also fails to interpret the absence of singing signal in *Delft* and *ZJU* (Bosschers 2018b).

On the other hand, Peng *et al.* (2017a) emphasized that vortex singing requires a closed sheet cavity at the tip and a twisted shape of the column cavity before singing. This suggests that either a helical mode wave is involved or there is an interaction between the sheet cavity and breathing waves propagating along TVC with the negative phase velocity at the zero-group-velocity condition. Determining these mechanisms requires more analysis of the dispersion relation in viscous flows.

In the exploration of vortex singing over the past thirty years, the greatest obstacle for both experiment and numerical simulation is the unclear triggering theory (Arndt 2002). Therefore, in this paper, the 2-D and 3-D resonance frequencies for TVC are proposed. Furthermore, the viscous corrections and analytical triggering criteria are given to explain how and where the TVC's singing should be triggered.

## 2. Theoretical analysis

### 2.1. *Resonance frequency of a vortex cavity in 2-D viscous flow*

In the study of resonance frequency for a vortex cavity in 2D viscous flow, the far-end of the TVC exhibits a fully developed interface, which can be characterized as a 2D axisymmetric cavitation bubble. Under isothermal conditions, the pressure within the bubble, $p_v$, is assumed constant. As the radius, $R$, of bubble increases, the bubble's wall displaces the surrounding liquid from the axis, leading to a decrease in tangential velocity due to the conservation of angular momentum and an increase of the local pressure, which resists the further inflation. Conversely, decreasing $R$ results in the increased tangential velocity and the decreased local pressure, preventing the bubble from collapsing. The associated resonance frequency, $f_n$, also known as the natural frequency of TVC is formulated as (Franc & Michel, 2016),

$$f_n = \frac{U_\infty}{2\pi r_c}\sqrt{\sigma K_\sigma \Big/ \ln\frac{r_\infty}{r_c}} \tag{2.1}$$

where $r_\infty$ is the distance from the tunnel's wall to the vortex axis, and $r_\infty >> R$ should be satisfied to ensure minimal impact on far-field pressure, $p_\infty$. Here, $r_c$ represents the equilibrium radius, $\sigma$ is the far-field cavitation number, which is defined by $(p_\infty - p_v)/(0.5\rho_L U_\infty^2)$, with $\rho_L$ and $U_\infty$ denoting the density and velocity of the liquid. $K_\sigma$ means the tangential stiffness coefficient.

The 2-D cavitation bubble oscillation is initially considered as the most probable resonance source of the vortex singing, because $f_n$ is triggered by the step change of the far field pressure but not affected by the step time (Osschers 2018b), therefore, accurate prediction of $f_n$ using Equation (2.1) is crucial. Conventional methods use $K_\sigma$ = 1.0 (Franc & Michel 2016) but overpredict the singing frequency due to the neglect of viscous effects (Bosschers 2009b). Determining $K_\sigma$ for a viscous cavitating vortex remains a challenge. In this paper, a new linearization strategy is physically proposed to resolve $f_n$.

Under the conditions of incompressible, zero mass transport across the bubble wall, neglecting the surface tension and non-condensable gas, the fluid velocity $u$ and pressure $p$ are determined by,

$$\frac{1}{r}(R\ddot{R}+\dot{R}^2-\frac{1}{r^2}R^2\dot{R}^2-u_\theta^2)=-\frac{1}{\rho_L}\frac{\partial p}{\partial r} \tag{2.2}$$

where $R$, $\dot{R}$ and $\ddot{R}$ are the radius, velocity and acceleration of bubble wall.

For the equilibrium condition, $\dot{R}$ = 0 and $\ddot{R}$ = 0, the Equation (2.2) is rewritten in the simplified form as,

$$-u_\theta^2/r=-\frac{1}{\rho_L}\frac{\partial p}{\partial r} \tag{2.3}$$

and in steady state, the azimuthal velocity $u_\theta$ follows the *Lamb-Oseen* vortex profile,

$$u_\theta(r)=\frac{\Gamma_\infty}{2\pi r}(1-e^{-a(r/r_{cv})^2}) \tag{2.4}$$

which $\Gamma_\infty$ is the circulation, $r_{cv}$ denotes the viscous core radius, and $a$ = 1.2564 helps $u_\theta$ to reach its peak at $r = r_{cv}$. Even though the *Lamb-Oseen* model does not precisely model the azimuthal velocity due to the lack of coupling with axial flows, it properly accounts for viscosity effects (Pennings 2016b). By integrating $u_\theta$ from bubble wall to far-field in the radial direction, the Equation (2.3) becomes,

$$\int_{r_c}^{r_\infty}\{[\frac{\Gamma_\infty}{2\pi r}(1-e^{-a(r/r_{cv})^2})]^2\Big/r\}dr=\int_{r_c}^{r_\infty}(\frac{1}{\rho_L}\frac{\partial p}{\partial r})dr \tag{2.5}$$

where the boundary condition $p(r = r_c) = p_v$ is used to solve the Equation (2.5),

$$\begin{aligned}&\frac{\rho_L\Gamma_\infty^2}{8\pi^2 r_{cv}^2}\{-\frac{r_{cv}^2}{r_\infty^2}-\frac{r_{cv}^2}{r_\infty^2}e^{-2a(r_\infty/r_{cv})^2}+\frac{2r_{cv}^2}{r_\infty^2}e^{-a(r_\infty/r_{cv})^2}-2aEi[-2a(r_\infty/r_{cv})^2]+2aEi[-a(r_\infty/r_{cv})^2]\\&+\frac{r_{cv}^2}{r_c^2}+\frac{r_{cv}^2}{r_c^2}e^{-2a(r_c/r_{cv})^2}-2\frac{r_{cv}^2}{r_c^2}e^{-a(r_c/r_{cv})^2}+2aEi[-2a(r_c/r_{cv})^2]-2aEi[-a(r_c/r_{cv})^2]\}\\&=p_\infty-p_v=\frac{1}{2}\rho_L U_\infty^2\sigma\end{aligned} \tag{2.6}$$

with $Ei$ corresponding to the exponential integral function.

Linearizing $r_c$ relative to $r_{cv}$ with a parameter $k_c$, i.e. $r_c = k_c r_{cv}$, the circulation, $\Gamma_\infty$, is derived from the Equation (2.6) by assuming $r_\infty >> r_{cv}$,

$$\Gamma_\infty^2 \approx \frac{4\pi^2 r_c^2 U_\infty^2 \sigma}{1+e^{-2ak_c^2}-2e^{-ak_c^2}+2ak_c^2 Ei(-2ak_c^2)-2ak_c^2 Ei(-ak_c^2)} \tag{2.7}$$

To describe the cavity oscillation dynamics, the Equation (2.2) with unsteady terms is solved using the time-varying vortex core radius $r_v$, substituting the Equation (2.4) into the Equation (2.2) and then integrating from $R$ to $r_\infty$ in the radial direction induces the equation,

$$R\ddot{R}\ln\frac{r_\infty}{R}+\dot{R}^2\ln\frac{r_\infty}{R}+\frac{1}{2}R^2\dot{R}^2\left(\frac{1}{r_\infty^2}-\frac{1}{R^2}\right)-\int_R^{r_\infty}\frac{\Gamma_\infty^2}{4\pi^2 r^3}(1-e^{-a(r/r_v)^2})^2\,dr=-\frac{p_\infty-p(R)}{\rho_L} \tag{2.8}$$

where the pressure on the bubble wall is calculated by,

$$p(R)=p_v-2\mu_L\frac{\dot{R}}{R} \tag{2.9}$$

here $\mu_L$ denotes the molecular viscosity of liquid. For small-amplitude oscillations, by the radial oscillation coefficient $k(t)$, $R = kr_{cv}$ is utilized to rewrite the Equation (2.8),

$$\begin{aligned}&\ddot{k}+\frac{2\nu_L}{r_{cv}^2k^2\ln\frac{r_\infty}{kr_{cv}}}\dot{k}+\left(\frac{1}{k}+\frac{r_{cv}^2}{2r_\infty^2}\frac{k}{\ln\frac{r_\infty}{kr_{cv}}}-\frac{1}{2k\ln\frac{r_\infty}{kr_{cv}}}\right)\dot{k}^2\\&+\frac{\Gamma_\infty^2}{8\pi^2r_{cv}^4}\left\{\frac{r_{cv}^2}{r_\infty^2}\frac{1}{k\ln\frac{r_\infty}{kr_{cv}}}+\frac{r_{cv}^2}{r_\infty^2}\frac{e^{-2a(r_\infty/r_v)^2}}{k\ln\frac{r_\infty}{kr_{cv}}}-\frac{2r_{cv}^2}{r_\infty^2}\frac{e^{-a(r_\infty/r_v)^2}}{k\ln\frac{r_\infty}{kr_{cv}}}+\frac{2aEi[-2a(r_\infty/r_v)^2]}{k\ln\frac{r_\infty}{kr_{cv}}}-\frac{2aEi[-a(r_\infty/r_v)^2]}{k\ln\frac{r_\infty}{kr_{cv}}}\right.\\&\left.-\frac{2aEi[-2ak^2(r_{cv}/r_v)^2]}{k\ln\frac{r_\infty}{kr_{cv}}}+\frac{2aEi[-ak^2(r_{cv}/r_v)^2]}{k\ln\frac{r_\infty}{kr_{cv}}}-\frac{1}{k^3\ln\frac{r_\infty}{kr_{cv}}}-\frac{e^{-2ak^2(r_{cv}/r_v)^2}}{k^3\ln\frac{r_\infty}{kr_{cv}}}+\frac{2e^{-ak^2(r_{cv}/r_v)^2}}{k^3\ln\frac{r_\infty}{kr_{cv}}}\right\}=\frac{-(p_\infty-p_v)}{\rho_L r_{cv}^2k\ln\frac{r_\infty}{kr_{cv}}}\end{aligned} \tag{2.10}$$

According to Choi *et al.* (2009), $k$ is assumed as constant when the singing cavity occurs with small-scale oscillations to linearize the Equation (2.10), i.e.

$$\frac{R}{r_v}=\frac{kr_{cv}}{r_v}\approx\frac{r_c}{r_{cv}}=k_c \tag{2.11}$$

By eliminating the nonlinear and high-order terms, and incorporating the Equation (2.6), the Equation (2.10) is simplified to a standard second-order form,

$$\begin{aligned}&\ddot{K}+2\xi\omega_n\dot{K}+\omega_n^2K=0\\&K=k-k_c\\&\xi\omega_n=\frac{\nu_L}{r_c^2\ln\frac{r_\infty}{r_c}}\\&f_n=\frac{\omega_n}{2\pi}=\frac{U_\infty}{2\pi r_c}\sqrt{\frac{\sigma}{\ln\frac{r_\infty}{r_c}}\frac{1+e^{-2ak_c^2}-2e^{-ak_c^2}}{1+e^{-2ak_c^2}-2e^{-ak_c^2}+2ak_c^2Ei(-2ak_c^2)-2ak_c^2Ei(-ak_c^2)}}\end{aligned} \tag{2.12}$$

where $K$ is the radial perturbation of $R$ relative to $r_{cv}$, $\xi$ is the damping ratio, and $\omega_n = 2\pi f_n$, represents the modified natural frequency.

Once $f_n$ is calculated by the Equation (2.12), the tangential stiffness coefficient $K_\sigma$ in the Equation (2.1) is updated to,

$$K_\sigma(k_c) = \frac{1+e^{-2ak_c^2} - 2e^{-ak_c^2}}{1+e^{-2ak_c^2} - 2e^{-ak_c^2} + 2ak_c^2 Ei(-2ak_c^2) - 2ak_c^2 Ei(-ak_c^2)} \tag{2.13}$$

Bosschers *et al.* (2018a) have pointed out the relations between cavitation inception number ($\sigma_i$) and $k_c = r_c/r_{cv}$. However, the onset of TVC is very sensitive to the nuclei of incoming flow, turbulence and scale effects (Zhang *et al.* 2015; Peng *et al.* 2017b), the desinent cavitation number ($\sigma_d$) is relatively stable and primarily influenced by the gas content (Arndt 2002; Song *et al.* 2017). Therefore, $\sigma_d$ is more suitable to reflect the relationship rather than $\sigma_i$. The solution of $\sigma_d$ is given by letting $r_c$ approach the vortex axis in the radial direction,

$$\begin{aligned} \sigma_d &= \left(p_\infty - p(r_c)\big|_{r_c\to 0}\right) \Big/ \frac{1}{2}\rho_L U_\infty^2 \\ &\approx \frac{\Gamma_\infty^2}{4\pi^2 U_\infty^2 r_{cv}^2} \left. \frac{1+e^{-2ak_c^2} - 2e^{-ak_c^2} + 2ak_c^2 Ei(-2ak_c^2) - 2ak_c^2 Ei(-ak_c^2)}{k_c^2} \right|_{k_c\to 0} = \frac{a \ln 2\, \Gamma_\infty^2}{2\pi^2 U_\infty^2 r_{cv}^2} \end{aligned} \tag{2.14}$$

By combining the Equations (2.7) and (2.14), $\Gamma_\infty$ can be eliminated, and the relative cavitation number, $\sigma/\sigma_d$, is obtained by,

$$\frac{\sigma}{\sigma_d} = \frac{1+e^{-2ak_c^2} - 2e^{-ak_c^2} + 2ak_c^2 Ei(-2ak_c^2) - 2ak_c^2 Ei(-ak_c^2)}{2a \ln 2\, k_c^2} \tag{2.15}$$

consequently, the tangential stiffness coefficient $K_\sigma$ is completely determined by the Equation (2.13) with $\sigma/\sigma_d$.

The experimental data measured from the stereo particle image velocimetry (SPIV) and high-speed video (HSV) for angles of attack equaling to 5°, 7° and 9° is provided by Bosschers (2018b) and marked in figures 1(a), (b) and (c). The desinent cavitation number, $\sigma_d$ is calculated by the Equation (2.14), where $r_{cv}$ has been proven to be equal to the vortex core radius in non-cavitating flows, therefore measured as 1.00*mm*, 1.15*mm* and 1.60*mm*, respectively. Note that the blue curves bound the error variation of 10% or 15% from the averaged value of cavity radius $r_{cv}$. The results indicate that all experimental data are covered by the predicted band of $r_c$ varying with $\sigma$, and the theoretical relation in the Equation (2.15) is verified.

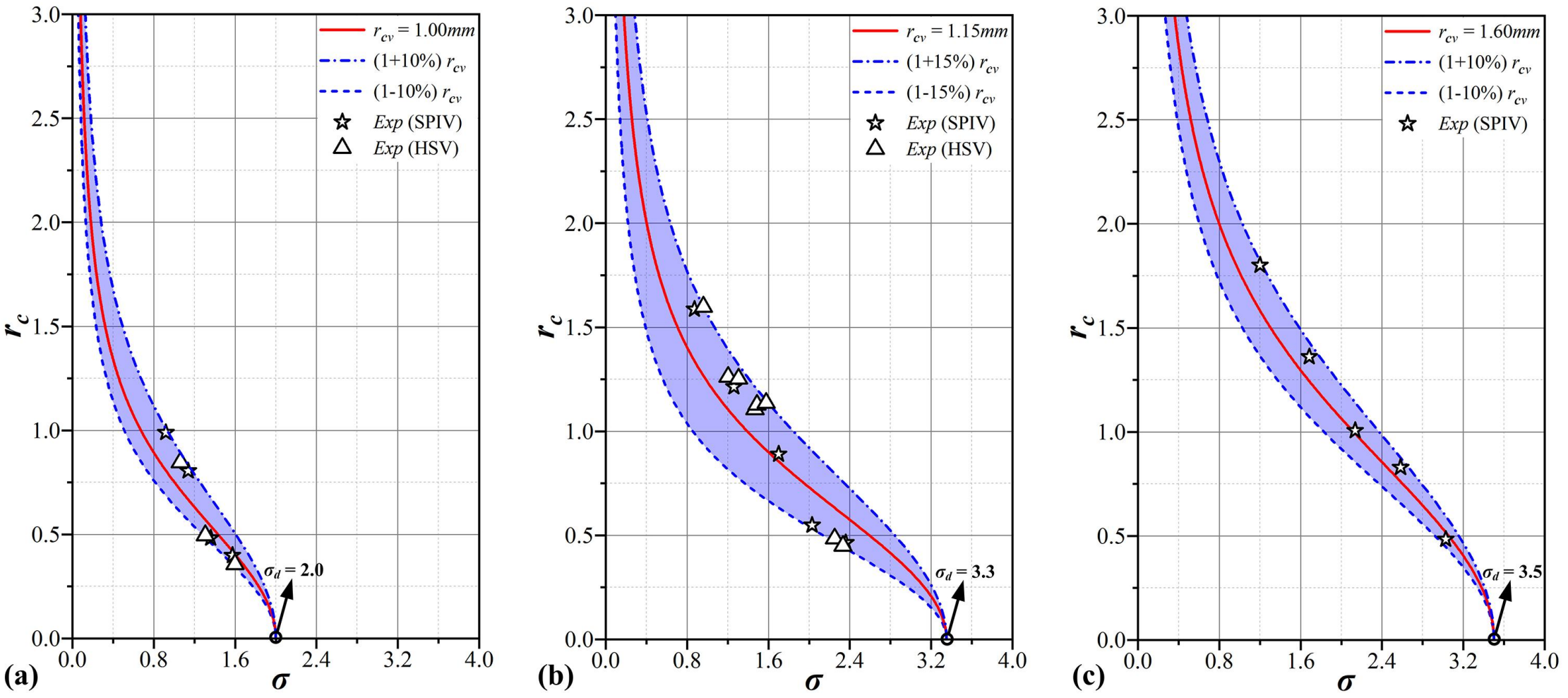

**FIGURE 1.** The measured cavity radius, $r_c$, varying with $\sigma$ obtained by SPIV and HSV for the angles of attack at 5°, 7° and 9°, the theoretical curves are depicted using the Equation (2.15) for (a) $r_{cv}$ = 1.00*mm* (with 10% error band), $\sigma_d$ = 2.0, (b) $r_{cv}$ = 1.15*mm* (with 15% error band), $\sigma_d$ = 3.3, and (c) $r_{cv}$ = 1.60*mm* (with 10% error band), $\sigma_d$ = 3.5.

Compared with semi-empirical models in Bosschers *et al.* (2018a), this approach contains no empirical parameters and provides an accurate fit of $r_c$ with $\sigma$. Therefore, the bubble's natural frequency in the Equation (2.1) is updated to,

$$f_n = \frac{U_\infty}{2\pi r_c}\sqrt{\sigma K_\sigma(\sigma/\sigma_d) \Big/ \ln\frac{r_\infty}{r_c}} \tag{2.16}$$

Based on above discussions, for a step change in far-field pressure, $f_n$ is believed to trigger the near-end TVC's resonance and singing if the incoming flow is properly changed to the singing condition. Note that although the Equation (2.16) has the same form as the Equation (2.1), $K_\sigma$ is completely different: $K_\sigma$ = 1.0 for the conventional method, and a function determined by $\sigma/\sigma_d$ for the proposed method.

### 2.2. *Dispersion relations for 3-D singing interfacial waves*

The theoretical analysis of self-excited oscillations for a 2-D cavitation bubble focuses on the triggering at the far-end of TVC. However, the key to understanding vortex singing lies in how those disturbances propagate upstream and further excite the resonance of the near-end cavity. Therefore, to explain why such humming tone is triggered only in a narrow range of cavitation number, the 3-D dispersion relations describing the cavity interfacial dynamics are utilized. According to Bosschers *et al.* (2018b), particularly for small-scale axial phase velocity or low frequency oscillation, the non-dimensional dispersion relation is expressed as,

$$\tilde{\omega}^{\pm}(\kappa,k_\theta) = \frac{2\pi f^{\pm} r_c}{U_\infty} = \frac{U_c}{U_\infty}\kappa + \frac{V_c}{U_\infty}[k_\theta \pm \sqrt{\frac{-\kappa K'_{k_\theta}(\kappa)}{K_{k_\theta}(\kappa)}}] \tag{2.17}$$

here $\tilde{\omega}$ is the resonant frequency, $\kappa = k_x r_c = -ik_r r_c$ denotes the wavenumber, and $k_\theta$ is the azimuthal wavenumber. $U_c$ and $V_c$ respectively stand for the mean axial and azimuthal velocities on the TVC surface, with $K_{k\theta}$ referring to a modified *Bessel* function of the second kind (Bosschers 2008).

The plus and minus signs in the Equation (2.17) correspond to two frequencies of each cavity oscillation mode. And surface tension effects are deemed negligible for a small cavity core size (Pennings *et al.* 2015a). Here figure 2 illustrates the primary cavity oscillation modes,

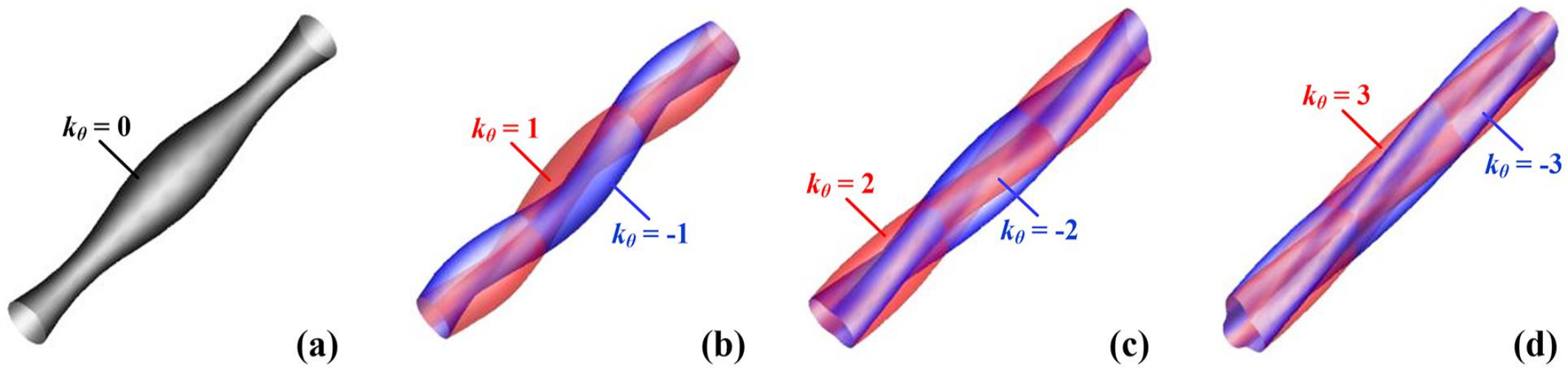

**FIGURE 2.** Shapes of cylindrical cavity oscillation modes for (a) monopole breathing mode at $k_\theta = 0$, (b) dipole centreline displacement mode at $k_\theta = \pm 1$, (c) quadrupole helical mode at $k_\theta = \pm 2$ and (d) hexapole helical mode at $k_\theta = \pm 3$.

Assuming $U_c / U_\infty = 1.0$ as per Bosschers (2008) due to the lack of analytical model for predicting $U_c$, we can derive the interfacial azimuthal velocity with the Bernoulli equation,

$$\frac{V_c}{U_\infty} = \sqrt{\frac{p_\infty - p_v}{0.5\rho_L U_\infty^{\ 2}}} = \sigma^{0.5} \tag{2.18}$$

The distributions of $U_c = U_\infty$ and $V_c = \sigma^{0.5} U_\infty$ adhere to the potential vortex solution. For a stationary TVC per the viscous *Lamb-Oseen* model, $V_c$ is replaced by the mean tangential velocity from Equation (2.4) when $r = r_c$, and Equation (2.18) becomes,

$$\frac{V_c}{U_\infty} = \frac{u_\theta(r_c)}{U_\infty} = \frac{\Gamma_\infty}{2\pi r_c U_\infty}[1 - e^{-a(r_c/r_{cv})^2}] \tag{2.19}$$

combining Equations (2.7), (2.13) and (2.15), the relation is further expressed as,

$$\frac{V_c}{U_\infty} = \sqrt{\sigma K_\sigma(\sigma/\sigma_d)} \tag{2.20}$$

To account for the axial acceleration effect induced by the reduction of tangential velocity, we apply the *Bernoulli* equation once more along the streamline developing from far-field to cavity surface, leading to,

$$\frac{U_c}{U_\infty} = \sqrt{\frac{p_\infty - p_v}{0.5\rho_L U_\infty^2} + 1 - (\frac{V_c}{U_\infty})^2} = \sqrt{\sigma + 1 - \sigma K_\sigma(\sigma/\sigma_d)} \tag{2.21}$$

Thus, the non-dimensional dispersion relation in Equation (2.17) for 3-D singing waves is recalculated using Equations (2.20) and (2.21),

$$\tilde{\omega}^\pm(\kappa, k_\theta) = \kappa\sqrt{\sigma + 1 - \sigma K_\sigma(\sigma/\sigma_d)} + \sqrt{\sigma K_\sigma(\sigma/\sigma_d)}[k_\theta \pm \sqrt{\frac{-\kappa K_{k_\theta}'(\kappa)}{K_{k_\theta}(\kappa)}}] \tag{2.22}$$

The revised equation for 3-D resonance frequencies is solely dependent on $\sigma$ and $\sigma/\sigma_d$ according to the Equation (2.22). Compared to the assumption of $U_c/U_\infty = 1.0$ in Bosschers (2008), Equation (2.22) is believed to improve the accuracy of predicting the wavelength of interfacial waves.

To summarize the three schemes for predicting the interfacial waves, the ratios of the tangential velocity, $V_c$, and axial velocity, $U_c$, to the incoming velocity, $U_\infty$, are presented in table 1, where scheme *I* is the conventional model using the potential vortex assumption, scheme *II* is the revised model by incorporating the *Lamb-Oseen* vortex model and scheme *III* is the newly proposed model in the present study. It is clear that the theoretical analysis with scheme *III* needs no any empirical parameter, and should be promising for predicting the complicated TVC dynamics. In practice, both the Equations (2.16) and (2.22) are much improved for predicting the resonance frequencies of 2-D cavitation bubble oscillations and the dispersion of 3-D TVC interfacial waves by incorporating the dependency on $\sigma/\sigma_d$.

**TABLE 1.** Two velocity ratios determined by different models.

| ***Scheme*** | ***I*** | ***II*** | ***III*** |
|---|---|---|---|
| $V_c/U_\infty$ | $\sigma^{0.5}$ | $(\sigma K_\sigma)^{0.5}$ | $(\sigma K_\sigma)^{0.5}$ |
| $U_c/U_\infty$ | 1.0 | 1.0 | $(\sigma+1-\sigma K_\sigma)^{0.5}$ |

## 3. Results and discussions

### 3.1. *Triggering for the vortex singing frequency*

The experiment conducted by *CSSRC*, operating at conditions of $U_\infty = 7m/s$ and a dissolved oxygen level of $D_O$ = 68%, serves to validate the equations presented in section **2**. According to *Peng et al.* (2017a), a twisted stationary TVC appears before singing, the noise levels abruptly intensify as the incoming pressure slowly increases until the cavitation number $\sigma$ reaches 1.40. By maintaining the pressure, the vortex continues to singing for several seconds or minutes due to the presence of resembling standing waves along the cavity surface, ceasing only as the vibration decays. The primary frequency of vortex singing identified as $f_s$ = 320*Hz* on the noise spectrum, aligns precisely with the vibration frequency observed from the 'dancing' cavity. The desinent cavitation number for this test is given as $\sigma_d \approx 2.30$ (Song *et al.* 2017), resulting in $\sigma/\sigma_d = 0.61$. The solved $f_n$ from Equation (2.16) is 319*Hz*, which closely matches the measured $f_s$ in the experiment. Further analysis, based on the revised dispersion relation in the Equation (2.22), reveals a zero-group-velocity point ($\partial\tilde{\omega}/\partial\kappa = 0$) located on the breathing mode curve ($k_\theta = 0^-$), yielding the frequency of $f_{zgv}$ = 324*Hz*, slightly larger than $f_s$. Although three frequencies $f_s$, $f_n$ and $f_{zgv}$ appear close quantitatively, their specific relationships require clearer elucidation.

The variations of cavity radius in time ($t$) and space ($x$), from two orthogonal views of the 'dancing' TVC, are recorded by the high-speed video and analyzed using *Canny*'s edge detection method (Canny 1986). As shown in figure 3(a), the *xy* plane view displays spindle-shaped radius contours distributed along the flow direction, representing helical waves with a fixed wavelength, $\lambda_s \approx 23.5mm$. Moreover, a laminar separation cavity (LSC) is observed attached to the tip of hydrofoil, anchoring the singing TVC. From the *xz* plane view in figure 3(b), cavity radius contours exhibit rhythmic vibration at the frequency $f_s$ in both time and space.

To quantify the cavity radius oscillations, $r_c$ is calculated as the mean cavity radius within the specified region between $x/C$ = 0.56 and $x/C$ = 1.56, yielding $r_c$ = 1.30*mm* with a pixel error of ± 0.18 *mm* on *xy* plane and ± 0.13 *mm* on *xz* plane. Using the Equation (2.21), $U_c/U_\infty$ is estimated to be 1.34. Notably, the slope of black solid line in figures 3(a) and (b) represents the convection flow velocity at $1.34U_\infty$. However, the slope of background contours exhibits an opposite trend to that of the convection line, suggesting that the singing waves are travelling from downstream toward the hydrofoil tip. This finding indicates that, the conjecture in earlier studies (Maines & Arndt 1997; Arndt *et al.* 2015) which speculated a standing wave along the surface of singing tip vortex appears to be a misunderstanding.

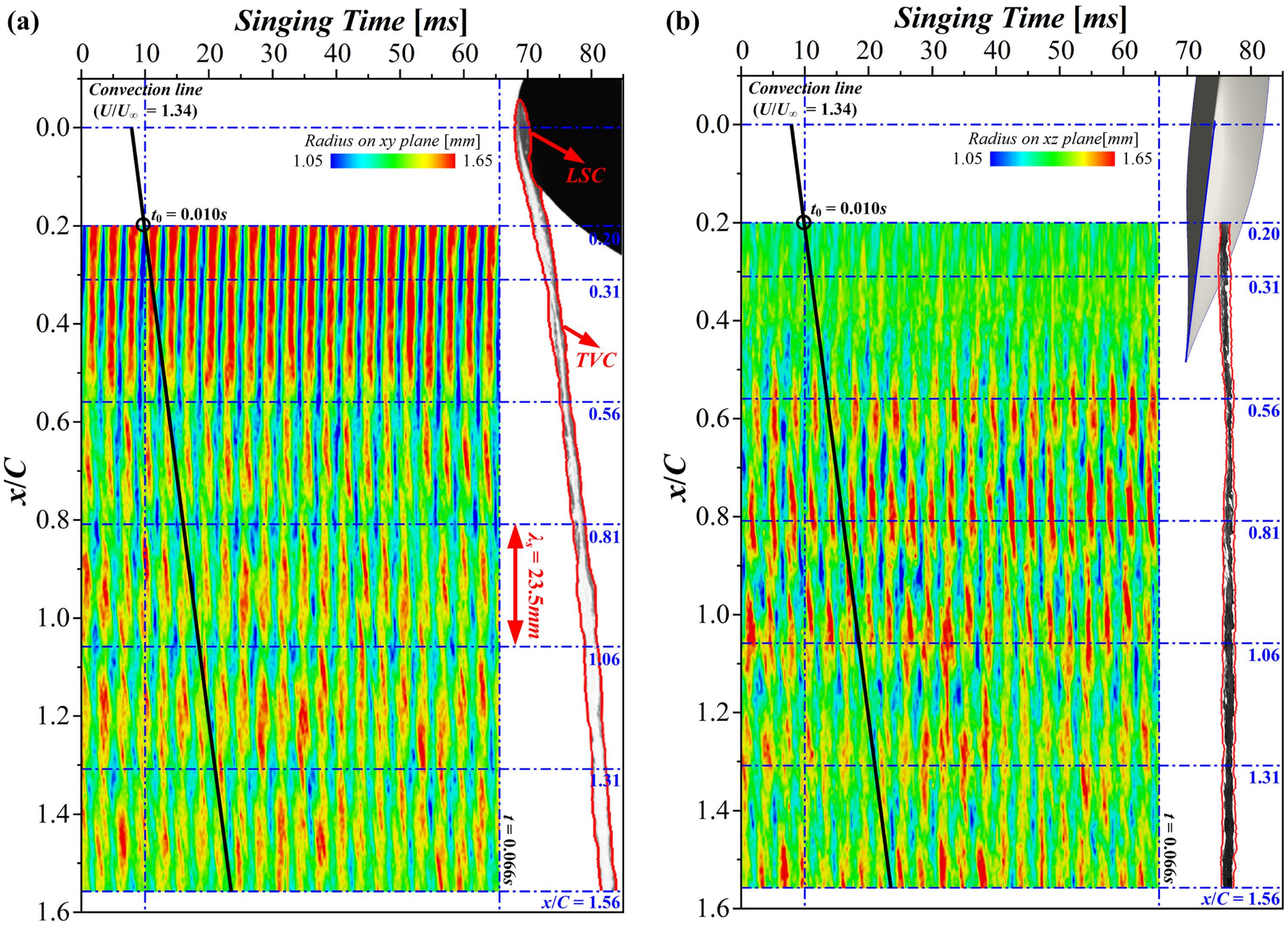

**FIGURE 3.** Variation of 'dancing' cavity radius in time and space with the (a) *xy* plane view and (b) *xz* plane view at the singing condition ($\sigma_s$ = 1.40, $\sigma_d \approx$ 2.30) provided by *CSSRC*, the cavity radius is identified by *Canny*'s edge detection method, and the black solid line indicates the interfacial convection flow at $1.34U_\infty$ predicted by the Equation (2.21).

The frequency-wavenumber diagram extracted from the *xy* plane view using a two-dimensional Fast Fourier Transform (2-D FFT) reveals interesting features. In figures 4, the non-dimensional frequency $\tilde{\omega} = 2\pi f r_c / U_\infty$ is defined by $r_c$, theoretical dispersion curves are predicted by different schemes discussed in table 1. The experimental data exhibit high-amplitude regions at specific frequencies, the first highlighted region situates at $\tilde{\omega}_s = -2\pi f_s r_c/U_\infty$ = -0.378, and the second highlighted region is located at the zero frequency, representing a stationary wave, associated with a singing wavenumber measured as $\kappa_s = 2\pi r_c/\lambda_s = 0.348$. It is also clear that there are two oblique bright stripes indicating high-amplitude cavity radius oscillations. As figure 4(a) indicates, none of the dispersion curves predicted by *Scheme I* aligns with the high-amplitude region. The frequency at zero-group-velocity point i.e. $f_{zgv}$ calculated by $\partial\tilde{\omega}/\partial\kappa = 0$ is plotted with the blue star mark, and is far away from the singing frequency, $\tilde{\omega}_s = -0.378$. For *Scheme II* as plotted in figure 4(b), two double helical modes i.e. $k_\theta = 2^-$ and $k_\theta = -2^+$ seem to approach the oblique bright stripes on the diagram within the first and fourth quadrants, with $f_{zgv}$ locating somewhat closer to $f_s$. However, the dispersion curve of $k_\theta = -2^+$ still exhibits considerable deviation from the right oblique bright stripe. For the results of *Scheme III* in figure 4(c), the dispersion curves' slope shows a better fitting accuracy of the highlighted background, and $f_{zgv}$ is much closer to $f_s$. The stationary wavenumber at the zero-frequency point, i.e. the wavenumber at the intersection of the dispersion curve of $k_\theta = -2^+$ and $\tilde{\omega} = 0$ is calculated as $\kappa_{zf} = 0.335$, which approaches the singing wavenumber $\kappa_s$. It is evident that *Scheme III* provides a more accurate representation of TVC dynamics under singing conditions compared to *Schemes I* and *II*.

From figure 4(c), it is noticed that the oscillation energy at the zero-group-velocity point manifests as a bright spot, located close to the intersection between dispersion curves of $k_\theta = -2^+$ and $k_\theta = 0^-$. However, further clarification is required to ascertain whether the most high-amplitude oscillation originates from the breathing mode of $k_\theta = 0^-$ or the double helical mode of $k_\theta = -2^+$.

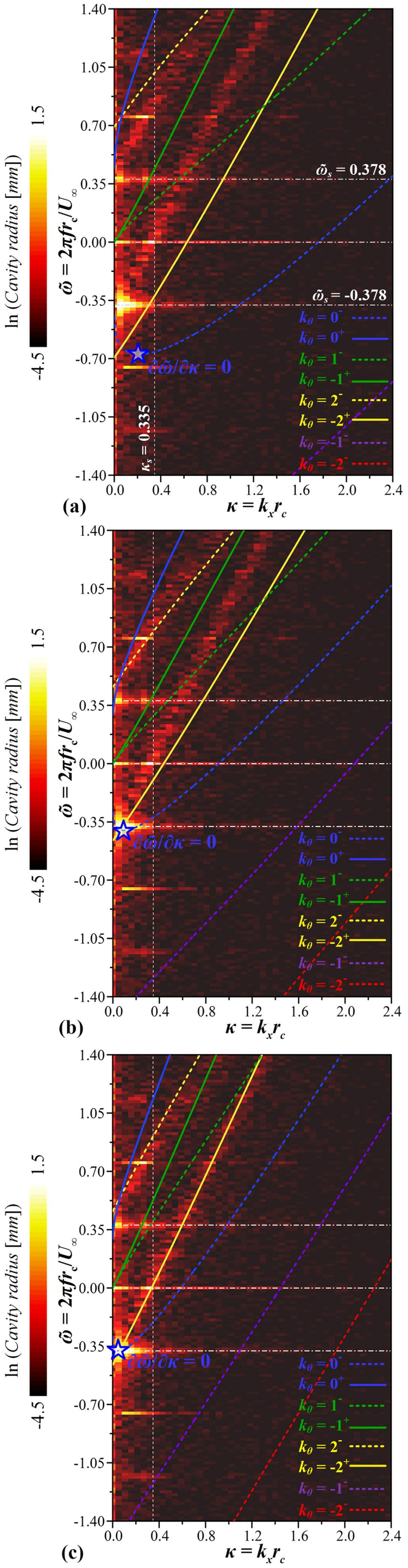

**FIGURE 4.** Frequency-wavenumber diagram of the variation of cavity radius on the *xy* plane at the singing condition ($\sigma_s$ = 1.40, $\sigma_d \approx$ 2.30) from *CSSRC*, the dispersion coefficients for comparison are obtained by (a) *Scheme I*, (b) *Scheme II* and (c) *Scheme III* (Iso colours indicate the cavity radius amplitude).

The frequency-wavenumber diagram from the *xz* plane view is illustrated in figure 5(a), where the first and second quadrants are plotted for validation according to the symmetry. Through *Scheme III*, the predicted dispersion curves match with the bright oblique stripes on the background, especially for the double helical waves of $k_\theta = 2^-$ and $k_\theta = -2^+$, while some faint bright stripes are also derived near the breathing mode curve of $k_\theta = 0^-$.

For further analysis, the coherence between two camera views of the *xy* plane and *xz* plane is presented in decibel and valued by,

$$Coherence(k_x, f) = \frac{G_{xy}(k_x, f) G_{xz}^*(k_x, f)}{\left|G_{xy}(k_x, f)\right| \left|G_{xz}(k_x, f)\right|} \tag{3.1}$$

where $G(k_x, f)$ is the 2-D FFT of cavity radius segments. The superscript, $^*$, represents complex conjugation.

Using the Equation (3.1), the phase difference spectrum is plotted in figure 5(b). At the zero-frequency point, the phase difference of double helical mode ($k_\theta = -2^+$) is approaching 180°. As for the breathing mode curve of $k_\theta = 0^-$, the phase difference is close to 0°, as expected. Moreover, the phase difference near the zero-group-velocity point is basically close to 0°, confirming that $f_{zgv}$ is stimulated by the breathing wave. However, from $f_s$ to $2f_s$, the phase difference for the helical wave of $k_\theta = -2^+$ is also near 0° due to its proximity to the breathing mode curve of $k_\theta = 0^+$. To summarize, once the near-end TVC starts to sing by the disturbance from the far-end cavity, the cavity interfacial distortions should be mainly composed of double helical waves and breathing waves that are located near the zero-group-velocity point.

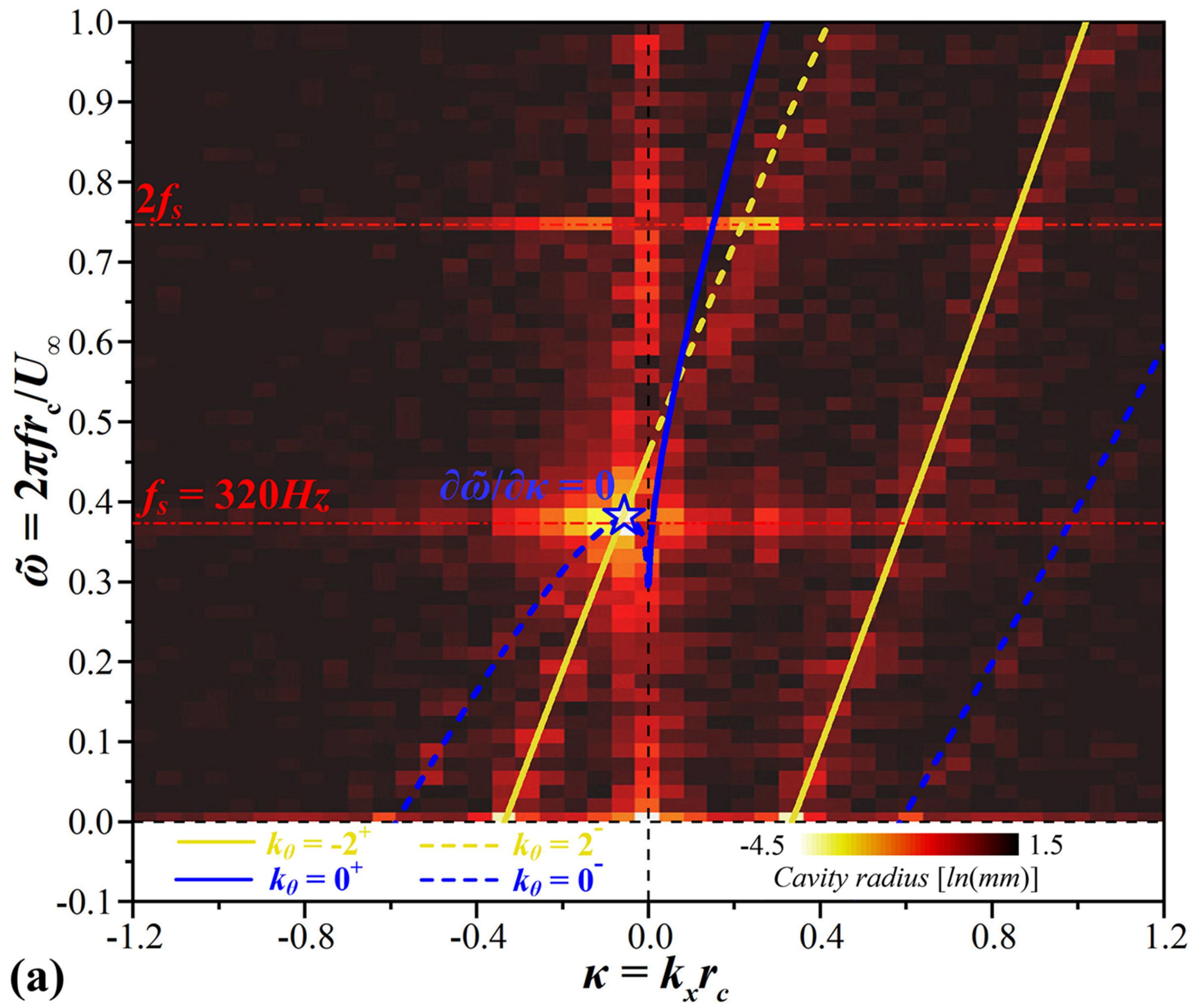

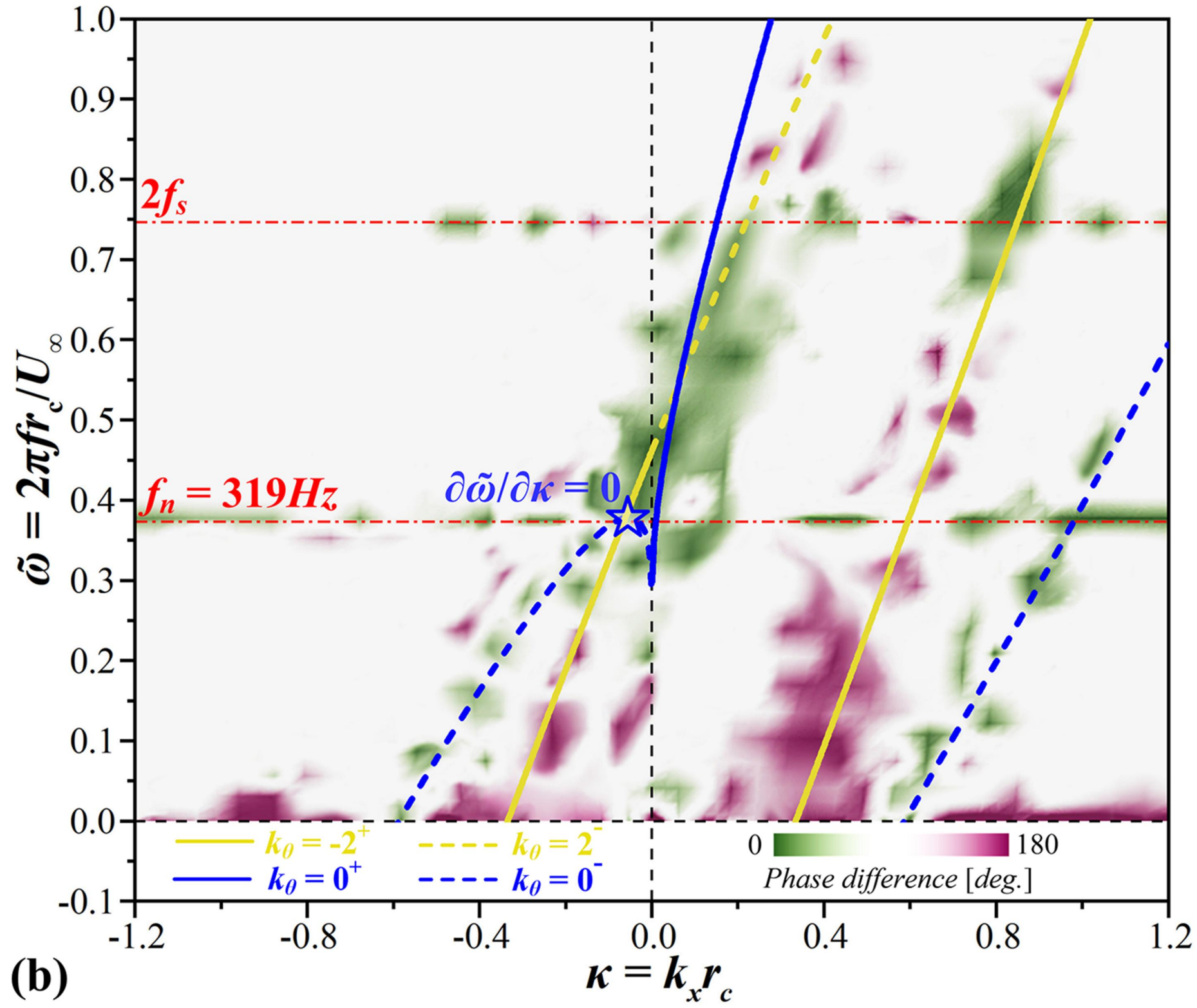

**FIGURE 5.** (a) Frequency-wavenumber diagram of cavity radius on the *xz* plane view at the singing condition ($\sigma_s$ = 1.40, $\sigma_d \approx$ 2.30) from *CSSRC*, the dispersion curves are obtained by *Scheme III*. (b) Phase difference between two camera views for the cavity radius.

To further reveal the triggering mechanism for a singing TVC, the cross-power spectral density (CPSD) of cavity radius variations from two orthogonal camera views is calculated by,

$$CPSD(k_x, f) = 120 + 10\log_{10}\{\frac{G_{xy}(k_x, f)G_{xz}^*(k_x, f)}{r_c^2}\} \tag{3.2}$$

As shown in figure 6(a), the local maxima of CPSD are concentrated firstly at the zero-group-velocity point on the breathing wave ($k_\theta = 0^-$) with frequency of $f_{zgv}$, and subsequently at the zero-frequency-point on the dispersion curves of $k_\theta = -2^+$ i.e. $f_{-2}{}^+ = 0$. Under this condition, the frequency, $f_i$, at the intersection between the dispersion curves at $k_\theta = -2^+$ and $k_\theta = 0^-$, marked with the circle, is almost equal to $f_{zgv}$ of 324$Hz$, indicating that the helical-shaped TVC starts to oscillate from a stationary state and then triggers $f_{zgv}$ successfully through the far-end oscillation at $f_n = 319Hz$.

In figure 6(b), the wavenumbers at the zero-frequency point $\kappa_{zf}$ and the zero-group-velocity-point $\kappa_{zgv}$ are solved as -0.335 and -0.054, respectively. Thus, the wavelength at the zero-frequency point i.e. $\lambda_{zf}$ is solved by $\lambda_{zf} = 2\pi r_c/\kappa_{zf} = -24.4mm$, and that at the zero-group-velocity-point i.e. $\lambda_{zgv}$ is also determined as $\lambda_{zgv} = 2\pi r_c/\kappa_{zgv} = -151.3mm$. denoting the relation of $\lambda_{zgv} \approx 6.2\lambda_{zf}$ for these two wavelengths. Furthermore, the phase velocity at the zero-group-velocity-point i.e. $V_{bw}$, calculated by $f_{zgv}\lambda_{zgv}$, is -49.0$m/s$, which is the slope of the purple solid line in figure 6(b). Note that the negative sign of phase velocity implies a breathing wave travelling upstream.

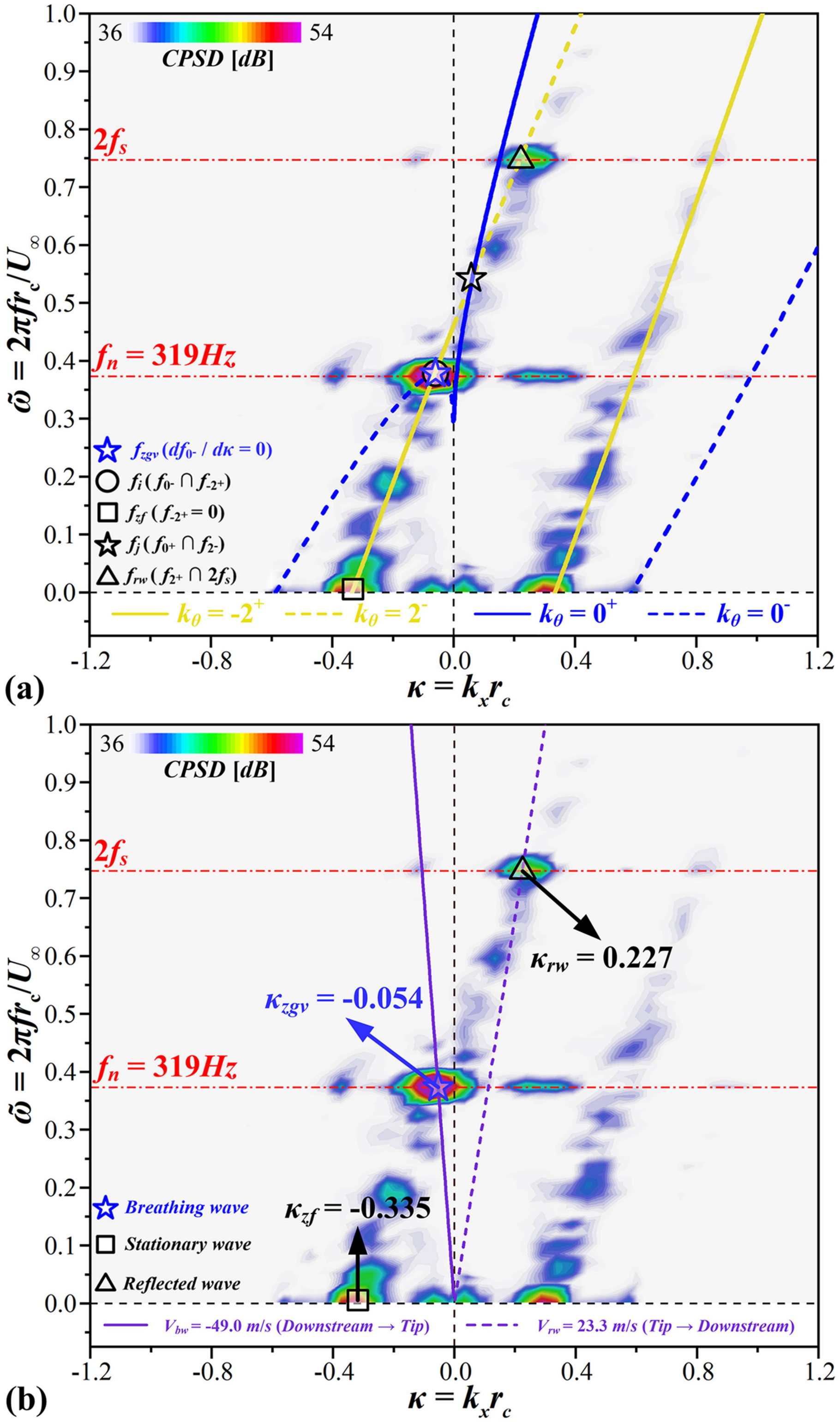

**FIGURE 6.** Frequency-wavenumber diagram of the cavity radius computed through *CPSD* at the singing condition ($\sigma_s$ = 1.40, $\sigma_d \approx$ 2.30) from *CSSRC*, with (a) theoretical feature points on dispersion curves deduced by *Scheme III* and (b) the phase velocity of breathing waves and reflected waves.

As figure 6(a) plots, there is also the region with a third highest amplitude of CPSD near the intersection between the line of $f_{rw}$ = 2$f_s$ and the dispersion curve of $k_\theta$ = $2^-$.

The subscript *rw* stands for a 'reflected wave' moving downstream, the wavenumber $\kappa_{rw}$ at the intersection is 0.227, and then the wavelength i.e. $\lambda_{rw}$ is calculated by $\lambda_{rw} = 2\pi r_c/\kappa_{rw} = 36.0mm$. Furthermore, the phase velocity $V_{rw}$ at the intersection is $23.3m/s$ calculated by $V_{rw} = 2f_s\lambda_{rw}$, which is the slope of the purple dash line in figures 6(b) and represents the speed of a downstream-propagating wave, implying that the breathing distortions seem to be basically absorbed by the sheet cavitation but generates limited reflected waves. Therefore, the oscillating LSC at the tip is not a 'trigger', as assumed by earlier studies, but more like a 'muffler', exerting little effect on the singing initiation but determining when the singing ends.

### 3.2. *Insights into singing vortex dynamics*

Except for the mean mode ($\tilde{\omega} = \kappa = 0$), waves at the zero-group-velocity and the zero-frequency points exhibit the two largest amplitudes of cavity radius oscillation, as illustrated in figures 4 and 5. To simplify the representation of these singing waves under *CSSRC*'s condition ($\sigma_s = 1.40$, $\sigma_d \approx 2.30$), $R_{zf}$ refers to the radius of helical wave ($k_\theta = -2^+$) at the zero-frequency point, the amplitude, frequency and wavelength are $A_{zf}$, $f_{zf}$ and $\lambda_{zf}$, respectively. $R_{zgv}$ is the radius of a breathing mode wave ($k_\theta = 0^-$) at the zero-group-velocity point, where the magnitude, frequency and wavelength are $A_{zgv}$, $f_{zgv}$ and $\lambda_{zgv}$ respectively. These parameters, in both two camera views for *xy* and *xz* planes are presented in figure 3, and given in table 2. It is evident that the wave amplitude at the zero-group-velocity, $A_{zgv}$, is larger than that at the zero-frequency point, $A_{zf}$, and the wavelength at the zero-group-velocity, $\lambda_{zgv}$, is much longer than that at the zero-frequency point, $\lambda_{zf}$, i.e. $\lambda_{zgv} \approx 6.2\lambda_{zf}$, as indicated before.

**TABLE 2.** The parameters applied for simplifying singing waves at the singing condition ($\sigma_s = 1.40$, $\sigma_d \approx 2.30$) from (1) the *xy* plane view and (2) the *xz* plane view.

| ***Parameters*** | $A_{zf}$ **[mm]** | $f_{zf}$ **[Hz]** | $\lambda_{zf}$ **[mm]** | $A_{zgv}$ **[mm]** | $f_{zgv}$ **[Hz]** | $\lambda_{zgv}$ **[mm]** |
|---|---|---|---|---|---|---|
| **(1) *xz plane*** | 0.19 | 0 | 24.4 | 0.32 | 324 | 151.3 |
| **(2) *xy plane*** | 0.21 | | | 0.36 | | |

To analyze features of a singing TVC, a synthetic wave $R_{zf+zgv}$ is constructed based on $R_{zf}$ and $R_{zgv}$, which represent two waves with the first two largest amplitudes of the cavity radius oscillation. Given the equilibrium radius of cavity, $r_c = 1.30mm$, the singing waves can be expressed as,

$$\begin{cases} R_{zf}(x,\theta,t) = r_c + A_{zf}\exp\{i(\frac{-2\pi x}{\lambda_{zf}} - 2\theta + 2\pi f_{zf}t)\} \\ R_{zgv}(x,\theta,t) = r_c + A_{zgv}\exp\{i(\frac{-2\pi x}{\lambda_{zgv}} + 2\pi f_{zgv}t)\} \\ R_{zf+zgv}(x,\theta,t) = r_c + A_{zgv}\exp\{i(\frac{-2\pi x}{\lambda_{zgv}} + 2\pi f_{zgv}t)\} + A_{zf}\exp\{i(\frac{-2\pi x}{\lambda_{zf}} - 2\theta + 2\pi f_{zf}t)\} \end{cases} \tag{3.3}$$

To validate this approach, we select four typical instants within the singing cycle. In figure 7, the cavity from the top view (*xz* plane) of *CSSRC*'s tunnel, showcasing the interfacial shapes with radii of $R_{zf}$, $R_{zgv}$ and $R_{zf+zgv}$ at each time instant. For each wave, the location with the maximum diameter is identified as the crest. The crest $A$, occurs at $x/C$ = 0.78 for the time instant $t_0$ = 0.010s. Then, the crest $A$, carried by breathing waves, propagates upstream from $x/C$ = 0.78 at $t_0$ to $x/C$ = 0.52 at $t_1$, which further approaches $x/C$ = 0.00 at $t_2$. At $t_2$, the crest $B$ is theoretically predicted at $x/C$ = 1.61. The distance between these two crests ($A$ and $B$) corresponds to $\lambda_{zgv}$, indicating that a large-scale volume pulsation is travelling along the singing TVC. By $t_3$ = 0.012$s$, crest $B$ reaches $x/C$ = 1.35 with $V_{bw}$ of -49.0m/s. The simplified singing waves, shaped by $R_{zf+zgv}$, faithfully reproduce the experimental details of a 'dancing' cavity.

From $t_2$ to $t_3$, both the upstream and downstream of the TVC undergo self-excited contraction and expansion, respectively. Guided by the traction-propulsion effect, the trough $C$ i.e. the location with the minimum diameter for each wave, along the $R_{zf+zgv}$ is forced to move upstream from $x/C$ =0.72 at $t_2$ to $x/C$ = 0.46 at $t_3$ with the phase velocity of $V_{bw}$. Therefore, the 2-D self-excited oscillations, occurring with a natural frequency, $f_n$ = 319$Hz$, as predicted by Equation (2.16), can trigger the wave at zero-group-velocity point and sustain 3-D breathing waves, which serve as the dominant source of vortex singing.

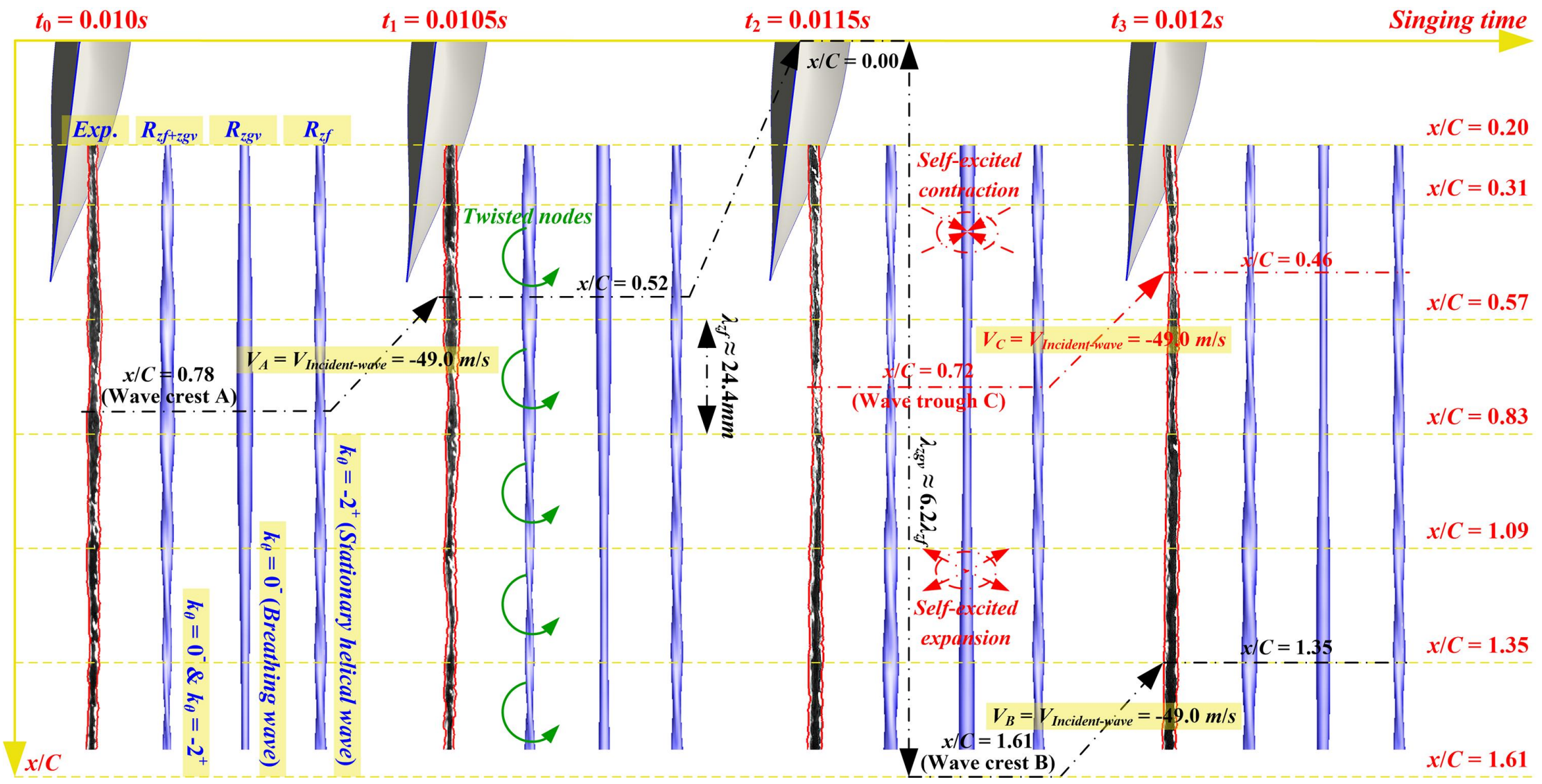

**FIGURE 7.** Four typical time instants within a singing cycle from the top view of *CSSRC*'s tunnel (*xz* plane), and the waves are simplified using the Equation (3.3) at the singing condition ($\sigma_s$ = 1.40, $\sigma_d \approx$ 2.30).

Considering the side view (*xy* plane) and utilizing the Equation (3.3), $R_{zf}$, $R_{zgv}$ and $R_{zf+zgv}$ are compared with experimental data in figure 8. The helical wave ($k_\theta$ = -2$^+$) exhibits a 180° phase difference between two views, e.g. the intersection between the yellow location line and the shape of $R_{zf}$ shifts from wave crests (maximum diameter) on the *xz* plane to wave nodes (minimum diameter) on the *xy* plane, and vice versa. Since there is no phase difference for breathing waves, the evolution of wave crests A and B, as well as the trough $C$ on the $R_{zgv}$ from $t_0$ to $t_3$ closely replicates that recorded from the top view. As figure 8 demonstrates, the synthetic wave $R_{zf+zgv}$ captures the strong volume pulsation travelling along the twisted TVC, with $\lambda_{zgv} \approx 6.2\lambda_{zf}$ keeping in good agreement with experimental observations. By $t_2$, the crest A approaches the tip ($x/C$ = 0.02) and begins to dissipate with the laminar separation cavity (LSC), leading to the periodic perturbation of tip separation bubbles. At $t_3$, the breathing wave is mostly absorbed by the attached LSC, with no observed reflected waves. Furthermore, under the traction-propulsion effects induced by TVC's self-excited oscillations, the crest B moves to $x/C$ =1.35 at $t_3$, which will initiate the next wave travelling towards the tip from downstream.

Learning from the 3-D cavity dynamics, we can conclude that the second high-amplitude region at the zero-frequency point shown in figure 4 denotes the stationary helical TVC with the wavelength approaching $\lambda_{zf}$. During vortex singing, the near-end wave with $f_{zgv}$ is fully triggered and sustained by the far-end 2-D resonance with $f_n$, inducing the in-phase breathing waves to propagate rapidly along the twisted TVC, thereby enhancing the far-field noise. These findings demonstrate that the earlier assumption attributing the singing TVC to standing waves is incorrect. Instead, the present results confirm that the singing TVC is determined by large-scale breathing waves travelling upstream.

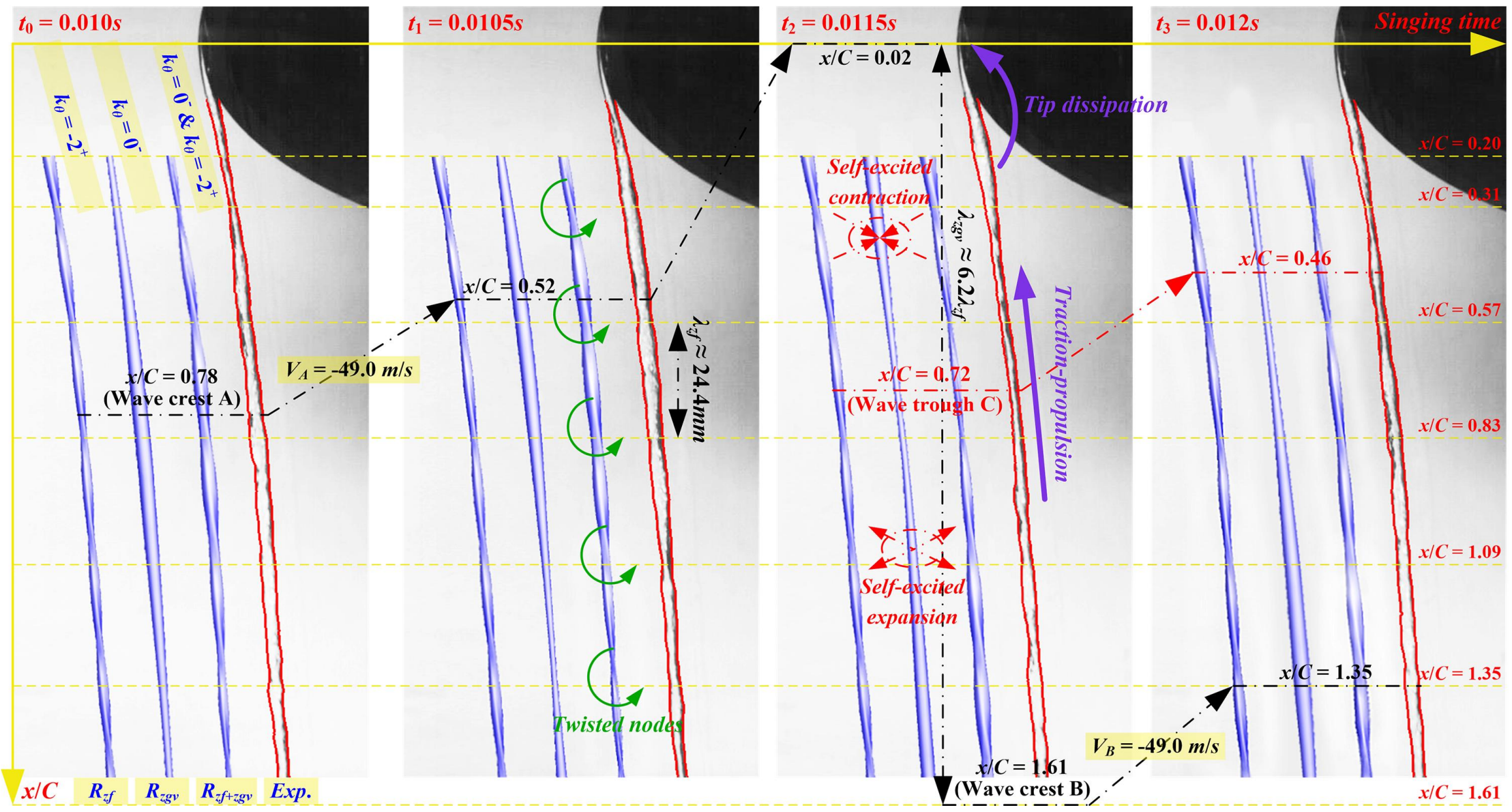

**FIGURE 8.** Four typical time instants within a singing cycle from the side view of *CSSRC*'s tunnel (*xy* plane), and the 3-D waves are simplified using the Equation (3.3) at the singing condition ($\sigma_s$ = 1.40, $\sigma_d \approx$ 2.30).

### 3.3. *Two triggering mechanisms for vortex singing*

The singing criterion has been preliminarily discussed in the section 3.2 based on cavity dynamics in *CSSRC*. Once the twisted far-end TVC begins to oscillate at the natural frequency $f_n$, which can be excited through a far-field pressure step, the cavity interfacial wave tends to disperse along $k_\theta = -2^+$ and further stimulates the oscillation at the zero-group-velocity point on $k_\theta = 0^-$. If the frequency at the intersection between the dispersion curves at $k_\theta = -2^+$ and $k_\theta = 0^-$, depicted by $f_i \sim \kappa$ ($f_0^- \cap f_{-2}^+$), sufficiently approaches the frequency at the zero-group-velocity, $f_{zgv} \sim \kappa$ ($df_0^-$ / $d\kappa$ = 0), the near-end cavity vibrations with $f_{zgv}$ should be triggered and further sustained. This process implies that $f_i = f_{zgv}$ is probably one of the singing criteria. By applying the Equation (2.22), the 3-D dimensionless resonance frequencies are given as,

$$\begin{cases} \tilde{\omega}^{-}(\kappa,0)=\dfrac{2\pi f_{0^{-}} r_c}{U_\infty}=\kappa\sqrt{\sigma+1-\sigma K_\sigma(\sigma/\sigma_d)}-\sqrt{\sigma K_\sigma(\sigma/\sigma_d)}\sqrt{\dfrac{-|\kappa|K_0'(|\kappa|)}{K_0(|\kappa|)}} \\ \tilde{\omega}^{+}(\kappa,-2)=\dfrac{2\pi f_{-2^{+}} r_c}{U_\infty}=\kappa\sqrt{\sigma+1-\sigma K_\sigma(\sigma/\sigma_d)}+\sqrt{\sigma K_\sigma(\sigma/\sigma_d)}\{-2+\sqrt{\dfrac{-|\kappa|K_{-2}'(|\kappa|)}{K_{-2}(|\kappa|)}}\} \end{cases} \tag{3.4}$$

here, if $\kappa > 0$, the wavenumber at the intersection $\kappa_i$ can be solved from the relation $\tilde{\omega}^{-}(\kappa, 0) = \tilde{\omega}^{+}(\kappa, -2)$, yielding $\kappa_i = 0.062$.

The singing criterion can be derived at the extremum by,

$$\left.\frac{\partial\tilde{\omega}^{-}(\kappa,0)}{\partial\kappa}\right|_{\kappa=\kappa_i}=\sqrt{\sigma+1-\sigma K_\sigma}-\frac{\sqrt{\sigma K_\sigma}}{2|\kappa_i|}\sqrt{\frac{-|\kappa_i|K_0'(|\kappa_i|)}{K_0(|\kappa_i|)}}\{1+\frac{|\kappa_i|K_0''(|\kappa_i|)}{K_0'(|\kappa_i|)}-\frac{|\kappa_i|K_0'(|\kappa_i|)}{K_0(|\kappa_i|)}\}=0 \tag{3.5}$$

by applying $\kappa_i = 0.062$, the following analytical relation can is achieved,

$$1+\frac{1}{\sigma}=3.445K_\sigma(\sigma/\sigma_d) \tag{3.6}$$

For convenience, the relation shown by Equation (3.6) is termed the first singing-line-*a*, representing one of triggering mechanisms for TVC-singing. As illustrated by Equations (2.13) and (2.15), $K_\sigma$ is solely dependent on $\sigma/\sigma_d$. Therefore, for each $\sigma/\sigma_d$, the theoretical singing cavitation number satisfying the Equation (3.6) can be defined, and the singing frequency within the second quadrant, $\tilde{\omega}^a$, is predicted using either of the Equations provided in (3.4),

$$\tilde{\omega}^a(\sigma/\sigma_d)=-\tilde{\omega}^{-}(\kappa_i,0)=0.488\sqrt{\frac{K_\sigma}{3.445K_\sigma-1}} \tag{3.7}$$

as $\sigma_d$ varies, the first singing-line-*a* is plotted by the blue curve in figure 9(a). Once $\sigma_d$ reaches 1.64, the curves with frequencies of $f_i$ and $f_{zgv}$ have the intersect at $\sigma/\sigma_d = 0.35$ marked with Triggered point-A, indicating that TVC-singing can only be triggered at the cavitation number of $\sigma = 0.35\times1.64 = 0.57$.

In figure 9(b), the 'dancing' cavity is expected to disperse along the curve at $k_\theta = -2^{+}$ and further stimulate the zero-group-velocity point with $\tilde{\omega} = 0.32$, allowing $f_{zgv}$ to be fully triggered through $f_i$ at the intersection.

However, as $\sigma_d$ increases, there are two solutions, $\kappa = -0.062$ and $\kappa = 0.062$, of $\sigma/\sigma_d = 0.35$, which satisfies the criterion $f_i = f_{zgv}$. According to Equations (3.6), (2.13) and (2.15), as $\sigma/\sigma_d$ approaches 0, then $k_c$ will approach $\infty$ and $K_\sigma$ will approach 1, resulting in $\sigma = 0.41$, which represents the minimum cavitation number for triggering vortex singing. Furthermore, if $\sigma$ is large enough, $K_\sigma$ approaches 0.29 and the solution of the Equation (3.6) is $\sigma/\sigma_d = 0.73$, where the asymptotic-line-*a* is located in figure 9(a).

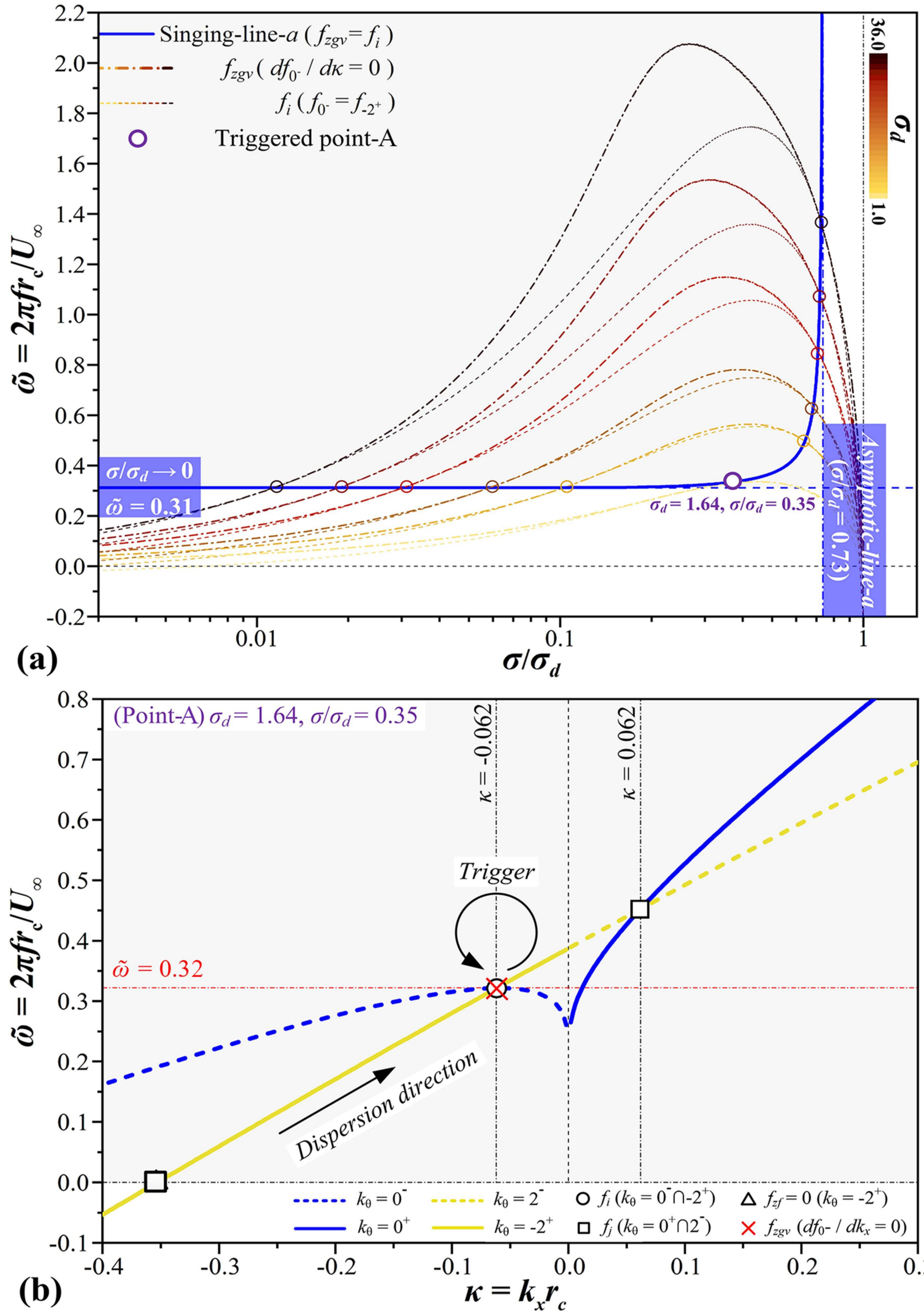

**FIGURE 9.** First type of vortex singing triggering, (a) theoretical solution for the frequency of lower singing-line-*a* varying with $\sigma/\sigma_d$, (b) first trigger mechanism at point A ($\sigma_d = 1.64$, $\sigma/\sigma_d = 0.35$).

In figure 10(a), once $\sigma_d$ much increases to 13.50, despite the large distance between $f_i$ and $f_{zgv}$, the frequency at the intersection between the dispersion lines on $k_\theta = 0^+$ and $k_\theta = 2^-$, i.e. $f_j \sim \kappa$ ($f_0^+ \cap f_2^-$), coincides with $f_{zgv}$ in the first quadrant as $\sigma/\sigma_d = 0.23$. The intersection $j$ marked with a block symbol in figure 10(b) has the same wavenumber of $\kappa_j = 0.062$ with the intersection $i$. Therefore, the interfacial waves on a static TVC disperse along $k_\theta = 2^-$, rather than $k_\theta = -2^+$, to excite the zero-group-velocity point with largest amplitudes of cavity radius oscillation at the singing condition. Consequently, $f_j = f_{zgv}$ is defined as the second criterion for TVC-singing, and the intersection $j$ is called Triggered point-B under the condition of $\sigma_d = 13.50$ and $\sigma/\sigma_d = 0.23$.

For $k_\theta$ = -$0^+$ and $k_\theta$ = $2^-$, dimensionless frequencies are derived from the Equation (2.22), i.e.

$$\begin{cases} \tilde{\omega}^+(\kappa,0)=\dfrac{2\pi f_{0^+} r_c}{U_\infty}=\kappa\sqrt{\sigma+1-\sigma K_\sigma(\sigma/\sigma_d)}+\sqrt{\sigma K_\sigma(\sigma/\sigma_d)}\sqrt{\dfrac{-|\kappa|K_0'(|\kappa|)}{K_0(|\kappa|)}} \\ \tilde{\omega}^-(\kappa,2)=\dfrac{2\pi f_{2^-} r_c}{U_\infty}=\kappa\sqrt{\sigma+1-\sigma K_\sigma(\sigma/\sigma_d)}+\sqrt{\sigma K_\sigma(\sigma/\sigma_d)}\left(2-\sqrt{\dfrac{-|\kappa|K_2'(|\kappa|)}{K_2(|\kappa|)}}\right) \end{cases} \tag{3.8}$$

if $\kappa > 0$, a constant of $\kappa_j$ = 0.062 is the solution to $\tilde{\omega}^+(\kappa, 0) = \tilde{\omega}^-(\kappa, 2)$. Solved from $\partial\tilde{\omega}_0^-/\partial\kappa = 0$, $\kappa_{zgv}(\sigma, K_\sigma)$ can be obtained as -0.38, inducing the singing tone at the zero-group-velocity point at $\tilde{\omega}$ = 1.03. Therefore, the second singing criterion is,

$$\tilde{\omega}^-(\kappa_j,2)=\tilde{\omega}^-(\kappa_{zgv},0) \tag{3.9}$$

As depicted in figure 10(a), if $\sigma_d$ increases continuing, the criterion of $f_j = f_{zgv}$ starts to present two solutions locating at both sides of $\sigma/\sigma_d$ = 0.23. These triggered points marked with circle symbol all behave in the second type and are further connected to the singing-line-*b*, shown as a red curve in figure 10(a) and satisfying the following relationship,

$$1+\frac{1}{\sigma}=1.422K_\sigma(\sigma/\sigma_d) \tag{3.10}$$

Similarly, if $\sigma/\sigma_d$ approaches 0, the minimum cavitation number can be calculated, i.e. $\sigma$ = 2.37. when $\sigma$ increases, $\sigma/\sigma_d$ approaches 0.41, where the asymptotic-line-*b* is obtained, as shown in figure 10(a). And the frequency $\tilde{\omega}^b$ is computed using either of formulas in Equations (3.8), e.g.

$$\tilde{\omega}^b(\sigma/\sigma_d)=\tilde{\omega}^-(\kappa_j,2)=0.625\sqrt{\frac{K_\sigma}{1.422K_\sigma-1}} \tag{3.11}$$

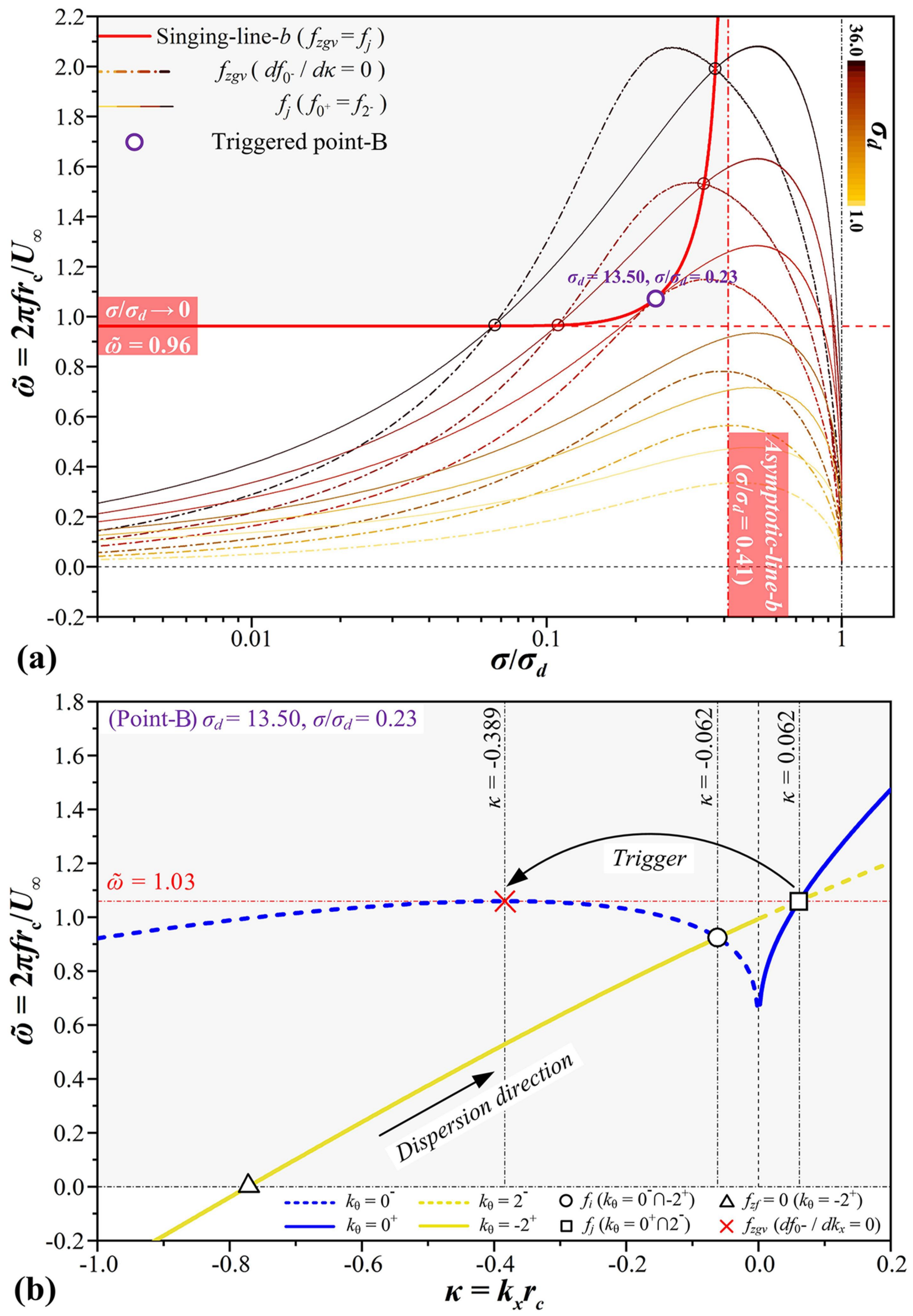

**FIGURE 10.** Second type of vortex singing triggering, (a) theoretical solution for the frequency of upper singing-line-$b$ varying with $\sigma/\sigma_d$, (b) second trigger mechanism at point B ($\sigma_d$ = 13.50, $\sigma/\sigma_d$ = 0.23).

To validate the two vortex singing lines, the dataset provided by *CSSRC* and *G.T.H.* is used for comparison on the diagram of ($\sigma^{0.5}$, $\tilde{\omega}$) as shown in figure 11. Note that the available cavity radius measured by *G.T.H.* is simply treated as $r_c$. For the singing-line $a$ representing the first criterion for the singing TVC, its non-dimensional frequency i.e. $\tilde{\omega}^a$ can be predicted by the relation of $\tilde{\omega}^a(\sigma) = 0.263\ (\sigma+1)^{0.5}$, which is obtained by combining the Equations (3.10) and (3.11). For the singing-line-$b$ representing the second criterion for the singing TVC, its non-dimensional frequency i.e. $\tilde{\omega}^b$ can be predicted by the relation of $\tilde{\omega}^b(\sigma) = 0.524\ (\sigma+1)^{0.5}$. The relations are described by two color solid lines, blue for the singing-line-*a*, and red for the singing-line-*b*.

As shown in figure 11, three sets of singing conditions in *CSSRC* under different dissolved oxygen contents are concentrated on singing-line-*a* and in the cavitation number range between $\sigma_{limit\text{-}a} = 0.41$ and $\sigma_{limit\text{-}b} = 2.37$. Here *A* and *B* represent the intersection points of $f_i \sim \sigma^{0.5}$ connecting $f_{zgv} \sim \sigma^{0.5}$ and $f_j \sim \sigma^{0.5}$ connecting $f_{zgv} \sim \sigma^{0.5}$, which are solved by the potential flow assumption and plotted with purple lines in figure 11. It is clear that in the case of $\sigma_{limit\text{-}a} \leq \sigma \leq \sigma_{limit\text{-}b}$, e.g. the singing region in *CSSRC*, the vortex singing can only be triggered along the singing-line-*a*. For the case of $\sigma > \sigma_{limit\text{-}b}$, e.g. four singing conditions in *G.T.H.*, the vortex singing is triggered along the singing-line-*b*.

In summary, we have presented and validated two theoretical vortex singing lines and explained why vortex singing can only occur within a narrow cavitation number range. This analysis provides insights into the underlying mechanisms governing the vortex singing and highlights the importance of hydrodynamic triggering conditions for such a whistling vortex cavity.

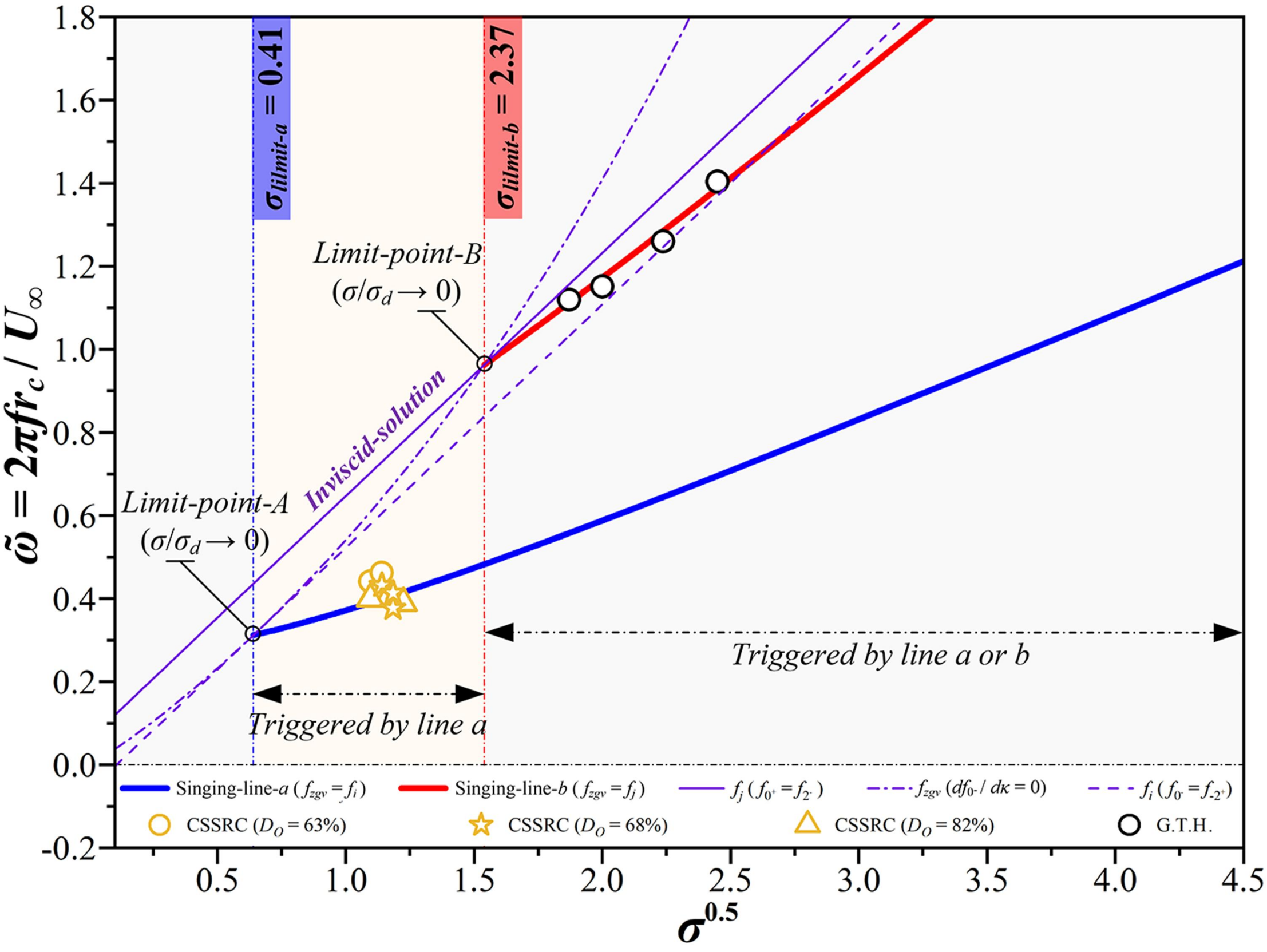

**FIGURE 11.** Theoretical solutions for the frequency of two vortex singing lines varying with $\sigma^{0.5}$, with singing conditions from *CSSRC* and *G.T.H.*

## 4. Conclusions

The phenomenon of vortex singing, a typical noise enhancement induced by the severe interfacial instability in tip vortex cavitating flows, has posed a significant challenge in cavitation hydrodynamic research over the past three decades. In this paper, theoretical analytical methods for predicting the 2-D resonance frequency of cavitation bubbles and the 3-D resonance frequency of cavity interfacial waves have been proposed without involving any empirical parameters.

(1) To accurately predict the resonance frequency of TVC, the equation for 2-D resonance frequency i.e. $f_n$ is established by introducing the *Lamb-Oseen* vortex model as well as the desinent cavitation number i.e. $\sigma_d$ to linearize the 2-D viscous cavitation bubble's dynamic equation, the dispersion model of 3-D cavity interfacial waves is improved by further updating the tangential and axial cavity interfacial velocity based on the *Bernoulli* equation.

(2) The wavenumber-frequency spectrum of the singing cavity radius, derived from the experiment by *CSSRC*, is convincingly replicated by the proposed 3-D dispersion model, indicating that singing waves predominantly consist of the stationary double helical mode ($k_\theta = 2^-$ and $-2^+$) and the breathing mode ($k_\theta = 0^-$), rather than standing waves, as assumed in previous literatures. It is also confirmed that far-end oscillations at $f_n$ play a crucial role in generating near-end breathing waves.

(3) Two trigger mechanisms have been proposed: The twisted TVC, initially at rest, is driven into motion by $f_n$ due to a step change of the far-field pressure. Subsequently, the frequency associated with the zero-group-velocity point ($f_{zgv}$) at $k_\theta = 0^-$ is excited through $f_i$, the frequency at the intersection of dispersion curves at $k_\theta = 0^-$ and $-2^+$, or $f_j$, the frequency at the intersection of dispersion curves at $k_\theta = 0^-$ and $2^-$, corresponding to the first and second type of triggering approach for vortex singing, respectively.

(4) By applying the cavitation number ($\sigma$) and relative cavitation number ($\sigma/\sigma_d$), solutions for vortex singing frequency ($f_s$) are determined independently for two type of trigger mechanisms, and are expressed by two triggering lines, which are validated by singing cases in *CSSRC* and *G.T.H.*, respectively.

(5) Analysis based on the coherence and the cross-power spectral density of cavity radius oscillation between two camera views for singing TVC depicts a negative and large phase velocity, indicating a large-scale breathing mode wave propagating along the cavity surface and travelling from downstream towards the hydrofoil tip.

**Acknowledgements:** This work was supported by the National Natural Science Foundation of China (Nos. 52336001, 52309117), and the China Postdoctoral Science Foundation (No: 2023M731895). The authors would like to acknowledge professor Y. Q. Liu from Shanghai Jiao Tong University and professor Q. Q. Ye from Zhejiang University for their contributions in the analytical study.

**Declaration of interests:** The authors report no conflict of interest.